\documentclass[journal]{IEEEtran}

\usepackage{setspace}
\usepackage{cite,graphicx,amsmath,amssymb}
\usepackage{subfigure}
\usepackage{url}
\usepackage{fancyhdr}
\usepackage{mdwmath}
\usepackage{mdwtab}
\usepackage{balance}
\usepackage{xcolor}
\usepackage{bm}
\usepackage{amsthm}
\usepackage{threeparttable}
\usepackage{algorithm}
\usepackage{algorithmic}
\usepackage{multirow}
\usepackage{flafter}
\usepackage{makecell}
\usepackage{diagbox}
\usepackage{booktabs}
\usepackage{multirow}
\usepackage{siunitx}

\usepackage{comment}

\usepackage{float}
\floatstyle{plaintop}
\restylefloat{table}
\usepackage{caption}

\providecommand{\url}[1]{#1}
\allowdisplaybreaks
\begin{document}
\title{CAPA: Continuous-Aperture Arrays for Revolutionizing 6G Wireless Communications}
\author{Yuanwei Liu, Chongjun Ouyang, Zhaolin Wang, Jiaqi Xu, Xidong Mu, and Zhiguo Ding\\
\thanks{Yuanwei Liu is with the Department of Electrical and Electronic Engineering, The University of Hong Kong, Hong Kong (email: yuanwei@hku.hk).}
\thanks{Chongjun Ouyang and Zhaolin Wang are with the School of Electronic Engineering and Computer Science, Queen Mary University of London, London, E1 4NS, U.K. (email: \{c.ouyang, zhaolin.wang\}@qmul.ac.uk).}
\thanks{Jiaqi Xu is with the Department of Electrical Engineering and Computer Science, University of California at Irvine, Irvine, CA 92697 USA (e-mail: xu.jiaqi@uci.edu).}
\thanks{Xidong Mu is with the Centre for Wireless Innovation (CWI), Queen's University Belfast, Belfast, BT3 9DT, U.K. (email: x.mu@qub.ac.uk).}
\thanks{Zhiguo Ding is with the Department of Computer and Information Engineering, Khalifa University, Abu Dhabi, United Arab Emirates, and also with the Department of Electronic and Electrical Engineering, The University of Manchester, M1 9BB Manchester, U.K. (e-mail: zhiguo.ding@ku.ac.ae).}
}

\maketitle
\begin{abstract}
In this paper, a novel continuous-aperture array (CAPA)-based wireless communication architecture is proposed, which relies on an electrically large aperture with a continuous current distribution. First, an existing prototype of CAPA is reviewed, followed by the potential benefits and key motivations for employing CAPAs in wireless communications. Then, three practical hardware implementation approaches for CAPAs are introduced based on electronic, optical, and acoustic materials. Furthermore, several beamforming approaches are proposed to optimize the continuous current distributions of CAPAs, which are fundamentally different from those used for conventional spatially discrete arrays (SPDAs). Numerical results are provided to demonstrate their key features in low complexity and near-optimality. Based on these proposed approaches, the performance gains of CAPAs over SPDAs are revealed in terms of channel capacity as well as diversity-multiplexing gains. Finally, several open research problems in CAPA are highlighted.
\end{abstract}

\section{Introduction}
The holy grail of multiple-antenna technology is to increase the number of antenna elements, thereby enhancing spatial degrees-of-freedom (DoFs) and achieving significant capacity gains in wireless networks. In recent years, several novel antenna array architectures have emerged, including reconfigurable intelligent surfaces \cite{di2020smart}, holographic multiple-input multiple-output (MIMO) \cite{deng2021reconfigurable}, and dynamic metasurface antennas \cite{shlezinger2021dynamic}. Despite their diverse architectures, these systems share a common evolutionary trend: \emph{larger} aperture sizes, \emph{denser} antenna configurations, \emph{higher} operating frequencies, and \emph{more flexible} structures. Naturally, this evolutionary trend leads to the creation of an (approximately) continuous electromagnetic (EM) aperture, i.e., a \emph{continuous-aperture array (CAPA)}, which is to be focused on in this paper.

\subsection{CAPA: Definition and Prototype}
Generally speaking, a CAPA operates as \emph{a single large-aperture antenna with a continuous current distribution}, which represents a significant shift from conventional spatially discrete arrays (SPDAs). It comprises a (virtually) infinite number of radiating elements coupled with electronic circuits and fed by a limited number of radio-frequency (RF) chains. A CAPA serves as an ideal transmission aperture by enabling continuous control over the amplitude and phase of the current over the surface of the CAPA where RF signals are modulated. It supports fully analog EM beamforming across the entire communication bandwidth, allowing precise alignment of the radiation pattern.

\begin{figure*}[!t]
\centering
    \subfigure[Kymeta Corporation's prototype (photo credit: Kymeta Corporation) {\copyright} CCBY \cite{sazegar2022full}. ``Rx'' and ``Tx'' denote receiver and transmitter, respectively.]
    {
        \includegraphics[height=0.32\textwidth]{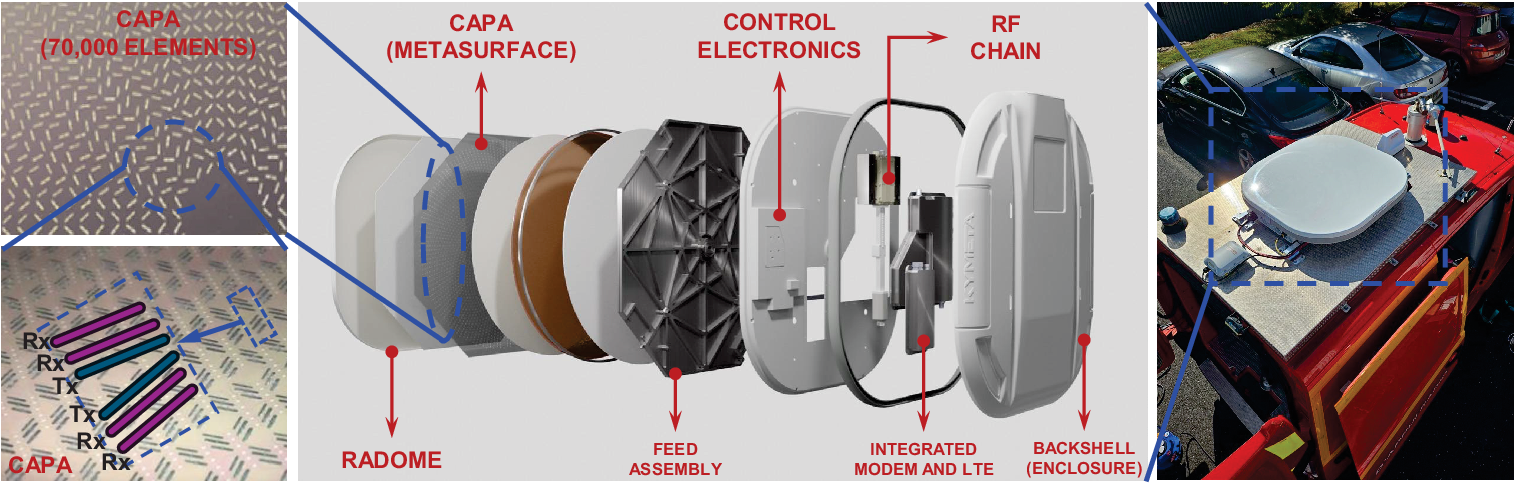}
	   \label{nc}	
    }\\
    \subfigure[MLWA (electrically-driven CAPA).]
    {
        \includegraphics[height=0.28\textwidth]{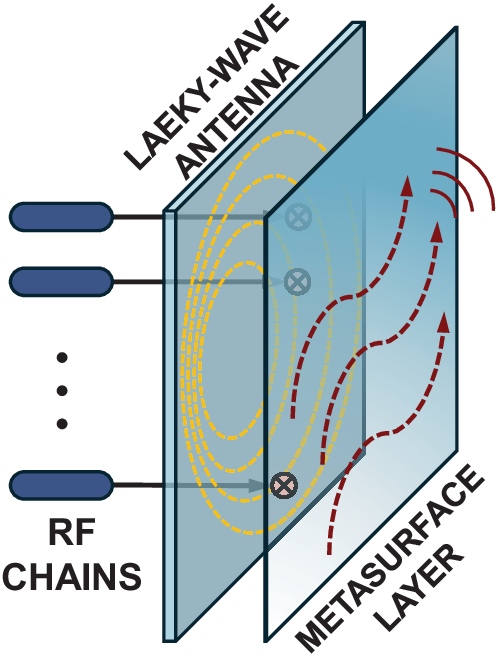}
	   \label{n0}	
    }
    \subfigure[OTCA (optically-driven CAPA).]
    {
        \includegraphics[height=0.28\textwidth]{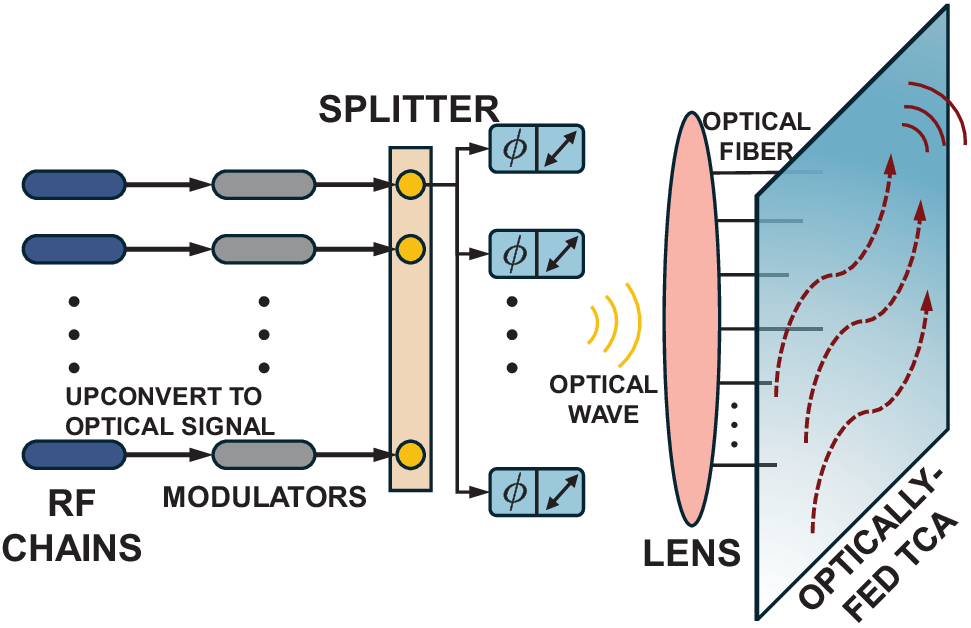}
	   \label{na}
    }
   \subfigure[ITGA (acoustically-driven CAPA).]
    {
        \includegraphics[height=0.28\textwidth]{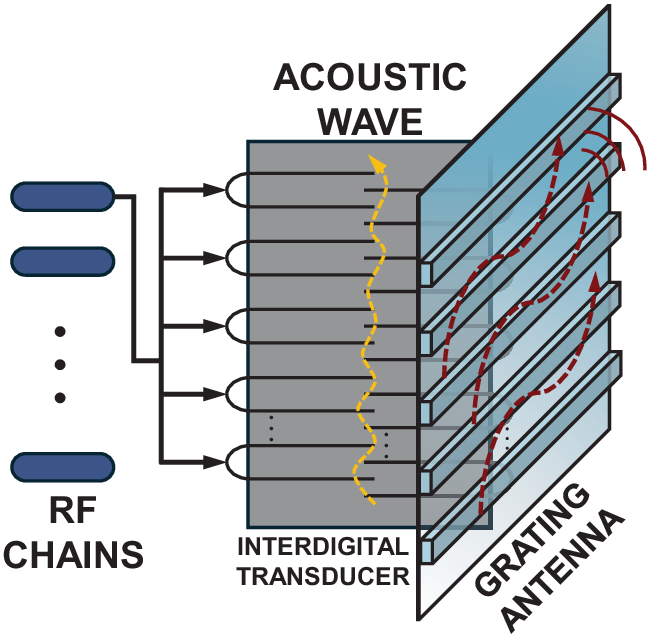}
	   \label{nb}	
    }
\caption{Illustration of a CAPA prototype and three types of CAPA hardware implementations, where the ``$\otimes$'' symbol in {\figurename} {\ref{n0}} represents an RF feed.}
\label{Fig_CAPA_imp}
\end{figure*}

The quest for continuous apertures in RF signal processing has been a longstanding goal in antenna design \cite{8000624}, with foundational contributions tracing back to the 1960s \cite{wheeler1965simple}. Recent advancements in materials and array fabrication have made the concept of CAPA feasible, leading to various CAPA prototypes \cite{8000624,black2017holographic,sazegar2022full}. Initial experimental studies indicate that these prototypes hold significant promise for improving system coverage and throughput in practical wireless network environments \cite{black2017holographic,sazegar2022full}. One CAPA prototype, developed by Kymeta Corporation, is illustrated in {\figurename} {\ref{nc}} \cite{sazegar2022full}. This design incorporates approximately 70,000 radiating elements, each equipped with a tunable capacitor, into a large circular aperture with an 82 cm diameter. The amplitude and phase of each element can be individually adjusted via software control. Owing to the ultra-dense deployment of these elements, the current distributions across the aperture can be conceptualized as continuous \cite{sazegar2022full}, effectively forming a CAPA.

\subsection{Key Advantages and Motivations}
By fully utilizing the whole surface of the aperture and enabling the free control of the current distribution, CAPAs can leverage spatial resources more \emph{effectively} and \emph{flexibly} than conventional SPDAs. This enables CAPAs to approach the theoretical performance limits. Therefore, CAPAs yield superior wireless transmission capabilities to SPDAs, making them a promising technology that can meet the challenging demands of sixth-generation (6G) networks. 

Furthermore, due to their continuous aperture, CAPAs rely on an \emph{integral-based continuous} signal model instead of a \emph{matrix-based discrete} one. This distinction introduces a novel framework for designing and analyzing extremely large-scale, ultra-dense arrays and offers the key benefits as follows:
\begin{itemize}
  \item \textbf{Enhanced Computation Efficiency:} Optimizing a current distribution, essentially a continuous function, can be computationally more efficient than optimizing high-dimensional matrices, which simplifies complex computations to a large extent.
  \item \textbf{Mathematical-Friendly Analysis:} Integrals are often easier to analyze and provide deeper insights than discrete summations. This aligns with the foundational motivations behind the invention of calculus by Leibniz.
  \item \textbf{Memory-Friendly Storage:} Storing a continuous function requires much less memory than storing a large matrix, leading to significant hardware storage savings.    
\end{itemize}

The above arguments highlight the significance of investigating CAPA applications for wireless communications, both in practice and theory, placing CAPAs at the forefront of the 6G wireless revolution. However, the design of CAPAs for wireless communications is still in its infancy. This motivates us to provide a systematic overview of CAPAs. In this article, we first review practical hardware implementations of CAPAs, followed by an exploration of their beamforming designs. We then analyze the performance gains of CAPAs over SPDAs. Finally, the open challenges of CAPAs are discussed to motivate future research.

\section{Practical Hardware Designs of CAPAs}
In this section, we present the practical hardware designs of CAPAs.
\subsection{Differences Between SPDAs and CAPAs}
An antenna is fundamentally a means of distributing EM currents across a carefully designed geometrical structure. The geometry determines how these currents interact with the surrounding EM field, shaping the radiated wave's directionality, polarization, and efficiency.

For SPDAs, reconfigurability is not implemented at the antenna level. This means the radiation pattern of each antenna in an SPDA is fixed once the antenna type is determined. Therefore, beams are entirely formed within digital or analog beamforming structures before RF signals enter the antenna. Additionally, the source currents in SPDAs are confined within each discrete antenna element, which significantly limits the diversity of source current distributions they can achieve. In contrast, CAPAs offer greater flexibility than SPDAs by enabling analog EM beamforming across the entire aperture and generating spatially continuous source current distributions.

We next introduce three key hardware implementations of CAPAs: electrically-driven, optically-driven, and acoustically-driven designs, as illustrated in {\figurename} {\ref{n0}}--{\ref{nb}}.
\subsection{Electrically-Driven CAPA}
Metasurface-based leaky-wave antennas (MLWAs) are a novel type of leaky-wave antennas (LWAs) that achieve amplitude and phase control of the current through an additional metasurface layer. As illustrated in {\figurename} {\ref{n0}}, an MLWA consists of an LWA with multiple RF feeds and a metasurface coating layer. 

The RF chain serves as a surface wave launcher for the LWA. When excited by the feed, a traveling wave propagates along the interface and decays exponentially into the surrounding medium. The traveling wave creates a continuous current distribution across the aperture, which radiates outward through the metasurface layer to free space. The metasurface layer, composed of resonant and scattering metamaterials, can be programmed to modulate the amplitude and phase of the wave at each point across the aperture. As a result, a modulated continuous current distribution is formed over the metasurface to achieve the desired radiation characteristics. 

Common control methods for metamaterials include lumped elements like PIN diodes, varactors, liquid crystals, and graphene. These MLWA designs are fabricated using flat-panel display technologies with tunable dielectric layers. Pivotal Commware \cite{black2017holographic} and Kymeta Corporation \cite{sazegar2022full} have recently commercialized these designs using lumped elements and liquid crystals, respectively.

The above process can be further summarized as follows:
\begin{enumerate}
    \item \textit{Feed Generation:} A surface-wave launcher generates a reference wave on the MLWA.
    \item \textit{Wave Propagation:} The reference wave travels either along a confined path, such as a microstrip line (waveguide mode) or across the surface (surface mode), yielding \emph{spatially-continuous} currents.
    \item \textit{Excitation and Modulation:} The propagated wave excites the metasurface. Controllers of the metasurface modulate the phase and amplitude of the excited currents.
    \item \textit{Radiation:} The \emph{modulated spatially-continuous} currents radiate from the metasurface to form the intended beam pattern.
\end{enumerate}

\begin{table*}[t!]
\centering
\caption{Summary of the Reviewed CAPA Implementations}
\label{Table: Compare_Hardware}
\resizebox{0.9\textwidth}{!}{
\begin{tabular}{|l|c|c|c|c|c|c|}
\hline
\textbf{Type} & \textbf{Signal Types$^\ddag$} & \textbf{Converter} & \textbf{Form Factor} & \textbf{Commercialization} & \textbf{Hardware Cost} & \textbf{Frequency Selection} \\
\hline
MLWA & RF signal & N/A & Large in size & Yes & Low & Moderate \\
\hline
OTCA & Optical signal & Photodiode & Small & No & High & Weak \\ 
\hline
ITGA & Acoustic signal & IDT & Compact & No &  Medium & Strong
\\ \hline
\multicolumn{7}{l}{$^\ddag${\footnotesize Here, we compare the different intermediate signals used in the structure. The final input and output signals are all in the RF domain.}}
\end{tabular}
}
\end{table*}
\subsection{Optically-Driven CAPA}
An optically driven tightly coupled array (OTCA) is realized using a tightly coupled array (TCA) of dipoles, each driven by a coherent optical feed network \cite{8000624}. As depicted in {\figurename} \ref{na}, an OTCA comprises multiple RF signal feeds, modulators that upconvert RF signals to optical frequencies, a splitter, optical phase and amplitude modulators, a lens, optical fibers, and a TCA. The array elements in the TCA are mutually connected to maintain a continuous current distribution \cite[{\figurename} 4]{8000624}.

Each dipole in the TCA is paired with a photodetector. These photodetectors convert continuous modulation of phases and amplitudes in the optical signals into the corresponding modulation in the transmitted RF wave. This enables precise analog control over phases and amplitudes, along with high-fidelity RF signal generation at each dipole, such that the entire array produces the desired beam radiation \cite{8000624}. As a result, the OTCA can closely approximate an ideal CAPA, with photodiodes serving as current sources for the coupled dipole elements.

The transmission process of an OTCA is outlined as follows:
\begin{enumerate}
    \item \textit{Signal Conversion:} The baseband digital signal is upconverted to the optical domain by modulating it using two phase-locked lasers, where the frequency difference between the lasers corresponds to the RF carrier frequency for array transmission \cite{8000624}. 
    \item \textit{Fiber Splitting and Routing:} The upconverted optical signals are split into multiple optical fibers, each routed to a unique optical phase/amplitude modulator and then a laser/transmitter.
    \item \textit{Optical Spatial Processing:} After obtaining the modulated optical signal, a lens is used to perform massive spatial processing---such as an analog Fourier transform---at the speed of light, in parallel, and with zero power consumption \cite{8000624}.
    \item \textit{Down-Conversion:} The optically processed signal is routed through optical fibers to photodetectors on the TCA. These photodetectors convert the optical signal back into the RF domain by exciting the tightly coupled dipoles, forming a \emph{continuous current distribution}.
\end{enumerate}
\subsection{Acoustically-Driven CAPA}
The key component of an interdigital transducer-based grating antenna (ITGA) is the interdigital transducer (IDT). IDTs convert electrical energy into mechanical (acoustic) waves via the piezoelectric effect and vice versa \cite{Interdigital}. The RF signal reconverted by the IDT is subsequently fed into a tightly coupled grating antenna array, where the elements are mutually connected to maintain a continuous current distribution \cite{Interdigital}. The slow propagation velocity of acoustic waves, such as surface acoustic waves on lithium niobate, allows sub-wavelength control of phase and amplitude, which is essential for synthesizing continuous apertures with precise radiation patterns \cite{morgan2010surface}.

As shown in {\figurename} {\ref{nb}}, IDTs consist of two comb-like electrode structures with ``fingers'' that are arranged in an alternating pattern, resembling interlocking fingers. Each electrode ``comb'' has multiple narrow, evenly spaced fingers, and the two combs interlace without touching each other. The transmitting process of an ITGA is described as follows:
\begin{enumerate}
    \item \textit{Signal Conversion:} The RF signal is converted into surface acoustic waves or bulk acoustic waves in the IDT through the piezoelectric effect.
    \item \textit{Acoustic Signal Processing:} After the RF signal is converted into acoustic waves, various signal processing operations, including filtering, frequency mixing, and phase-shifting, can be done through engineered finger geometries \cite{morgan2010surface}. 
    \item \textit{Back Conversion and Radiation:} Multiple IDTs act as receiving transducers, reconverting the acoustic waves back into RF signals. These RF signals are then fed into a grating antenna array, forming a continuous current distribution for radiation.
\end{enumerate}
\subsection{Comparison of Different Implementations}
Table {\ref{Table: Compare_Hardware}} summarizes the characteristics of the three reviewed CAPA implementations. The key difference between these implementations is the types of analog signals used during signal processing within the CAPA structures. For MLWAs, analog beamforming is performed directly on upconverted RF signals, similar to conventional RF transmitters. In contrast, OTCAs and ITGAs convert the upconverted RF signals into optical and acoustic waves, respectively. 

Since different signal types have different wavelengths, the form factors of these CAPA implementations can differ significantly. For example, consider an EM wave and a surface acoustic wave both operating at $1$ GHz. The EM wave has a wavelength of approximately $0.3$ m, while the acoustic wave has a wavelength of several micrometers. Therefore, ITGAs can be made more compact compared to MLWAs. For OTCAs that use optical signals, the size of the implementation depends on the optical wavelength used.

Another crucial aspect is the frequency response of these implementations. Among the three, OTCAs allow optical signals to propagate freely around the lens, while MLWAs and ITGAs confine signals within waveguides or IDT filters, where only specific frequencies propagate efficiently. As a result, ITGAs exhibit the strongest frequency selectivity, followed by MLWAs, while OTCAs have the weakest frequency selectivity. Due to this characteristic, ITGAs and MLWAs may require multiple separate CAPA structures to support different subcarrier frequencies, whereas OTCAs can handle multi-carrier signals with a single CAPA.

\section{Beamforming Designs for CAPAs}
A primary challenge in CAPA systems is designing the source current distribution across the surface, i.e., designing the analog EM beamforming. In contrast to SPDAs, the EM channel and beamformer for CAPAs are represented as continuous functions rather than discrete vectors or matrices. In this context, the utility function (e.g., spectral efficiency) that we aim to maximize through EM beamforming design becomes a \emph{functional}---a specialized type of function that accepts other functions as inputs and produces a scalar output. Consequently, conventional design approaches based on discrete optimization are not directly applicable to CAPA beamformer design. We will introduce several promising approaches to address this challenge in the following.

\subsection{Discretization Approach}
The most straightforward approach is to approximate continuous channels by discretizing them into finite-dimensional vectors or matrices, thereby only the corresponding discrete beamformers need to be designed. A common method for achieving this in wireless systems is the Fourier transform \cite{sanguinetti2022wavenumber}. Specifically, using the Fourier transform, the spatial-domain representation of the CAPA channel can be converted into the wavenumber-domain representation, which exhibits several favorable properties for discretization. First, the wavenumber-domain channel is band-limited, with its dominant values confined within an elliptical region. Second, half-wavelength sampling in the wavenumber domain closely approximates the true values with negligible errors. As a consequence, the continuous functional design problem for CAPA beamforming can be closely approximated by a discrete problem in the wavenumber domain. Therefore, the conventional discrete optimization method and the existing optimization toolboxes can be applied. It is worth noting that this discretization approach in the wavenumber domain is not limited to design problems with specific utility functions, making it a versatile approach for CAPA beamforming. However, a significant challenge with this approach is its potentially high computational complexity because the number of samples required to accurately approximate the true channel increases exponentially with the signal frequency and the CAPA aperture size, resulting in high-dimensional variables that need to be optimized and unaffordable computational complexity.

\begin{figure*}[!t]
\centering
    \subfigure[Illustration of the heuristic beamforming design for CAPAs, where the MRC/MRT beamformer aligns with the user's spatial response, and the ZF beamformer is orthogonal to the interference subspace.]
    {
        \includegraphics[width=0.35\textwidth]{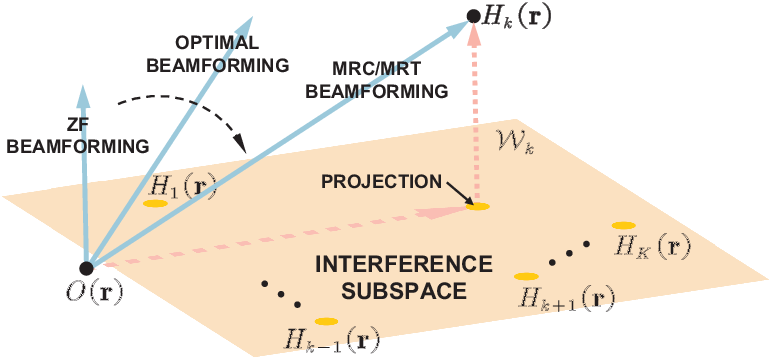}
	   \label{fig_Optimal}	
    }\hspace{5pt}
    \subfigure[Illustration of the subspace approach, where $J_k(\mathbf{r})$ represents the beamformer for user $k$. Any arbitrary $J_k(\mathbf{r})$ can be decomposed into two components: one within $\mathcal{W}$ and the other orthogonal to $\mathcal{W}$.]
    {
        \includegraphics[width=0.35\textwidth]{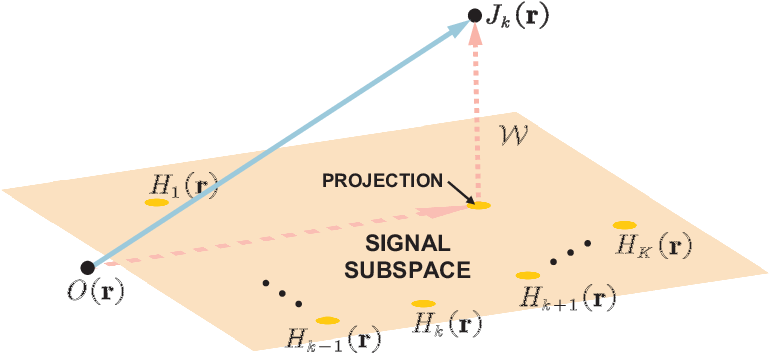}
	   \label{fig_Subspace}	
    }\hspace{5pt}
    \subfigure[Spectral efficiency achieved by different approaches in a downlink narrowband system with a uni-polarized CAPA transmitter and four uni-polarized single-antenna receivers, operating at 2.4 GHz. The impact of EM mutual coupling is omitted for the sake of brevity. Other simulation parameters can be found in \cite{wang2024beamforming}.]
    {
        \includegraphics[width=0.35\textwidth]{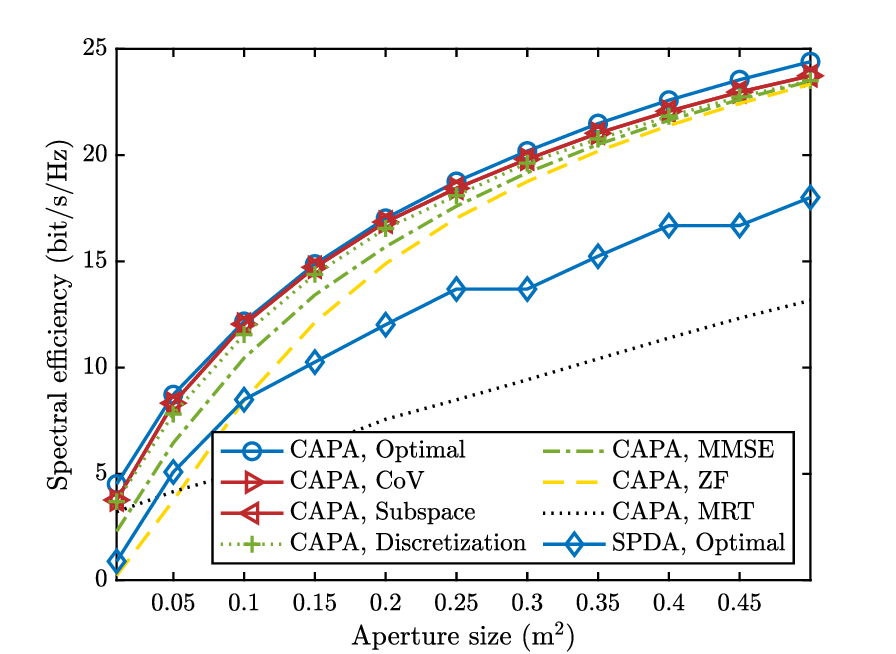}
	   \label{fig_SE}
    }
\caption{Illustration of CAPA beamforming. For {\figurename} {\ref{fig_Optimal}} to {\ref{fig_Subspace}}, a CAPA is used to serve $K$ uni-polarized single-antenna users, where $H_k({\mathbf{r}})$ represents the spatial response of user $k$, with $\mathbf{r}$ denoting a point within the CAPA's aperture. Here, $\mathcal{W}$ denotes the signal subspace spanned by the spatial responses of all users, while $\mathcal{W}_k$ represents the interference subspace for user $k$, spanned by the spatial responses of all users except user $k$. Additionally, $O({\mathbf{r}})$ marks the origin of the corresponding signal subspace.}
\end{figure*}

\subsection{Calculus of Variations Approach}
The calculus of variations (CoV) is an effective approach for identifying the optimal function that achieves the extremum of a functional. Variations extend the concept of differentiation. In conventional discrete beamforming optimization, the first-order derivative (defined through differentiation) of the objective function (for unconstrained problems) or the Lagrangian function (for constrained problems) with respect to the discrete optimization variables plays a crucial role in finding the optimal discrete beamformers. Similarly, using the concept of variations, the first-order derivative of the corresponding continuous functional with respect to the continuous input functions can be defined, providing a necessary condition for determining the optimal continuous beamformers, which can be used for algorithm designs. The CoV approach enables the direct design of continuous CAPA beamformers without requiring discrete approximations. Thus, compared to the discretization approach, it often achieves better performance, eliminates the necessity of handling high-dimensional vectors or matrices, and provides more insights into the structure of the optimal solution.

However, the CoV approach may not be directly applied to CAPA beamforming design involving complex utility functions and constraints, as the associated first-order derivative conditions can be too complicated to optimize in these cases. In such scenarios, heuristic methods offer an alternative by aiming to design closed-form CAPA beamformers that are near-optimal in most cases and fully optimal in specific cases, thereby eliminating the need for complex optimization. This is achieved by carefully selecting parameters within the optimal beamformer structure obtained through the CoV. For conventional SPDAs, closed-form designs such as maximum-ratio transmission/combination (MRT/MRC), zero-forcing (ZF), and minimum mean-squared error (MMSE) beamforming have been developed based on the optimal structure. These heuristic methods can be extended to CAPA systems, with their expressions theoretically derived in \cite{ouyang2024performance} for uni-polarized single-antenna users, as depicted in {\figurename} {\ref{fig_Optimal}}.

\subsection{Subspace Approach}\label{Secton: Subspace Method}
An alternative approach is to leverage the property that the optimal CAPA beamformers reside within the subspace spanned by a specific set of known basis functions, where those components orthogonal to the subspace have no impact on system performance \cite{ouyang2024performance}. Thus, the optimal CAPA beamformers can be expressed as a linear combination of these basis functions, requiring only the design of the coefficients in this linear combination \cite{ouyang2024performance}. In this case, the functional optimization problem with respect to the continuous CAPA beamformer is converted into an optimization problem with respect to the discrete coefficients, allowing the use of the conventional discrete optimization method. A critical step in the subspace approach is to identify an appropriate set of basis functions, which is closely linked to the form of the utility function and constraints, and typically requires deep analysis of the system under consideration. For example, it has been shown that optimal CAPA beamformers for maximizing communication spectral efficiency reside in the subspace spanned by the continuous channels of each user, as illustrated in {\figurename} {\ref{fig_Subspace}}. While the subspace approach is highly efficient once a suitable set of basis functions is established, it often lacks insight into the structure of the optimal solutions that can be important for practical low-complexity designs.

\begin{table*}[t!]
    \centering
    \caption{Comparison of Different Approaches for CAPA Beamforming}
    \label{Table: Compare_Beamforming}
    \begin{tabular}{|c|c|c|}
    \hline
    \textbf{Approach} & \textbf{Advantage} & \textbf{Disadvantage} \\ \hline
    Discretization & \makecell[l]{{\small$\bullet$} High generality \\ {\small$\bullet$} Discrete optimization} & \makecell[l]{{\small$\bullet$} High computational complexity \\ {\small$\bullet$} Limited insights} \\ \hline
    Calculus of Variations & \makecell[l]{{\small$\bullet$} Low computational complexity \\ {\small$\bullet$} Insights into optimal solutions} & \makecell[l]{{\small$\bullet$} Requires new algorithms \\ {\small$\bullet$} Relative low generality } \\ \hline
    Subspace & \makecell[l]{{\small$\bullet$} Low computational complexity \\ {\small$\bullet$} Discrete optimization} & \makecell[l]{{\small$\bullet$} Requires suitable subspace basis \\ {\small$\bullet$} Limited insights} \\ \hline
    \end{tabular}
\end{table*}

In Table {\ref{Table: Compare_Beamforming}}, we compare the advantages and disadvantages of different approaches for CAPA beamforming.

\subsection{Case Study}
{\figurename} \ref{fig_SE} provides a numerical example to compare the spectral efficiency achieved by the aforementioned approaches. Here, we assume an ideal CAPA capable of fully supporting analog EM beamforming. The impact of different hardware implementations can be further explored by introducing specific constraints on the source current. As a baseline, the optimal solution of the formulated problem is included, which is obtained using the highly complex monotonic optimization method \cite{ouyang2024performance}. It is observed that the CoV and subspace approaches achieve identical performance, closely approaching the optimal solution, while the discretization approach incurs a slight performance loss. This is expected, as the first two approaches directly solve the original CAPA beamforming problem, whereas the discretization approach leads to an approximated solution. The heuristic MMSE design exhibits performance with acceptable loss in most cases, which is promising in practice. Furthermore, regardless of the design approaches, CAPAs significantly enhance the spectral efficiency compared to SPDAs, with performance gains increasing as the aperture size grows. This result validates the capability of CAPA to fully exploit spatial DoFs. 

\section{Performance Analysis of CAPAs}
We now analyze CAPAs' performance gains over SPDAs, where the used performance evaluation metrics include channel capacity, diversity gains, and multiplexing gains.
\subsection{Channel Capacity}
\subsubsection{Single-User Channel Capacity}\label{Section: Single-User Channel Capacity}
This section analyzes the capacity of single-user channels between two CAPAs. For SPDA-based systems, the MIMO channel is represented as a matrix that can be decomposed into parallel, non-interfering single-input single-output sub-channels using singular value decomposition (SVD). These sub-channels enable optimal power allocation through a water-filling strategy to achieve the channel capacity.

Characterizing the single-user capacity of CAPAs is particularly challenging due to the continuous nature of their spatial responses, which are modeled by an \emph{integral operator} rather than a discrete matrix. In physically realistic CAPA-based channels, the spatial response operator (radiation operator) is analytic and square-integrable. This compactness allows the operator to be represented using a complete orthogonal basis set for both transmit currents and received fields. Through \emph{Hilbert-Schmidt decomposition (or SVD)}, the operator can be expressed as a weighted sum-product of these bases.

\begin{figure*}[!t]
\centering
    \subfigure[Single-user capacity for different propagation distances (shown in the parenthesis).]
    {
        \includegraphics[height=0.23\textwidth]{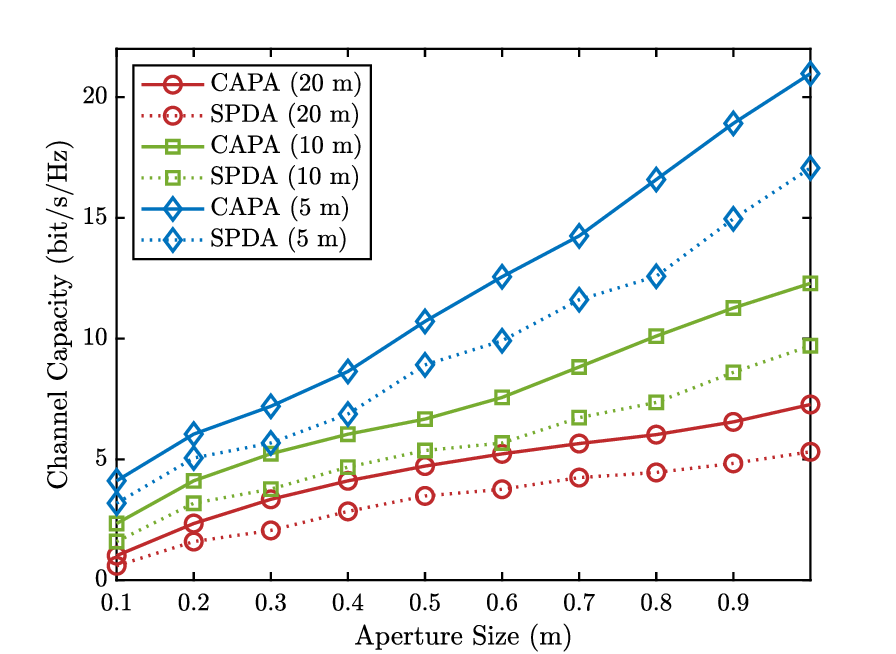}
	   \label{figSectionIV_Figure_1a}	
    }
    \subfigure[Two-user MAC capacity region for different aperture sizes (shown in the parenthesis).]
    {
        \includegraphics[height=0.23\textwidth]{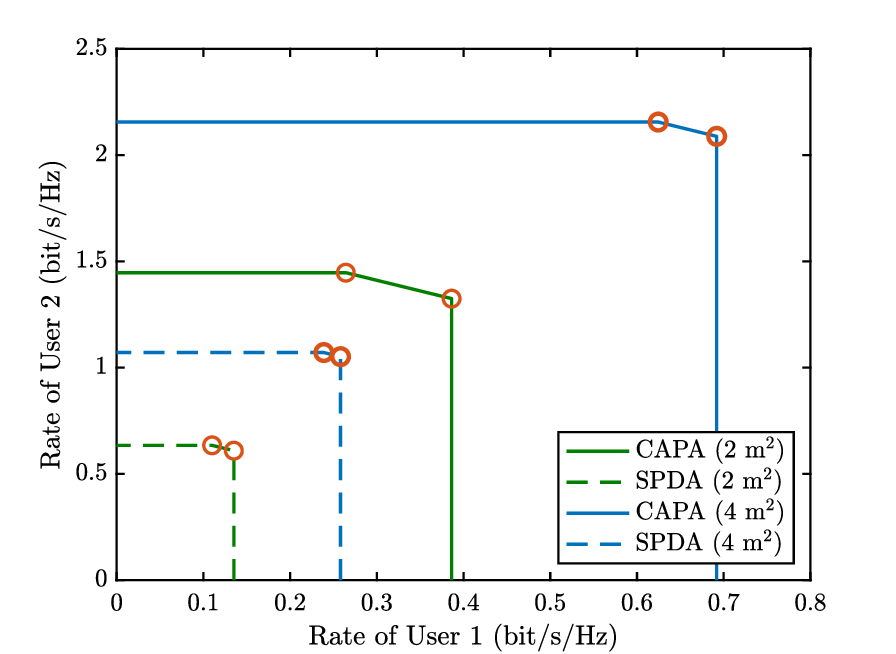}
	   \label{figSectionIV_Figure_1b}	
    }
   \subfigure[Two-user BC capacity region for different aperture sizes (shown in the parenthesis).]
    {
        \includegraphics[height=0.23\textwidth]{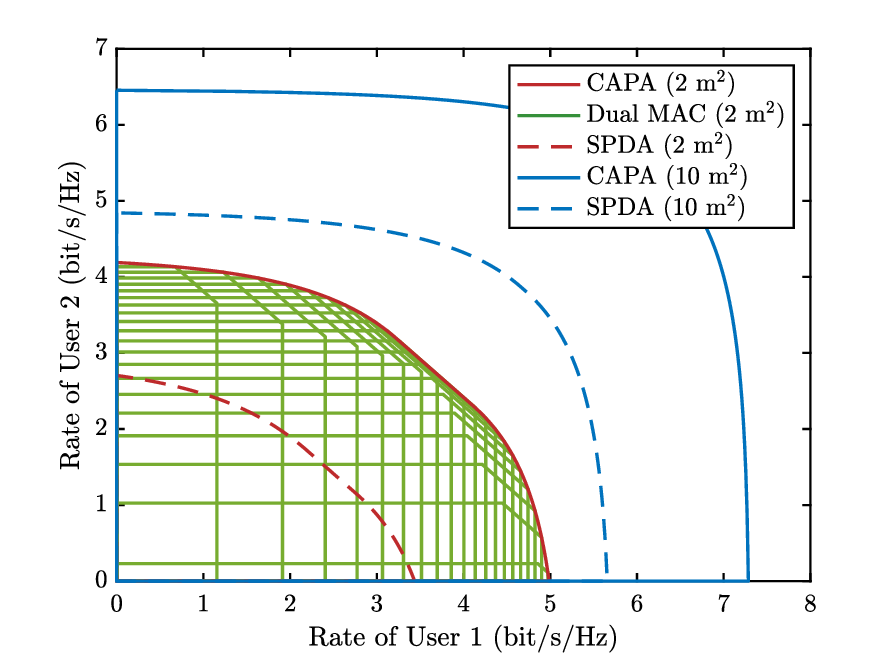}
	   \label{figSectionIV_Figure_1c}	
    }
\caption{The single-user capacity and the multiuser capacity region of CAPA-based LoS channels. For the channel capacity shown in {\figurename} {\ref{figSectionIV_Figure_1a}}, both the transmit CAPA and the receive CAPA are configured as linear arrays, with additional simulation parameters outlined in \cite{sanguinetti2022wavenumber}. For the two-user MAC and BC capacity regions shown in {\figurename} {\ref{figSectionIV_Figure_1b}} and {\figurename} {\ref{figSectionIV_Figure_1c}}, respectively, the simulation parameters are detailed in \cite{zhao2024continuous}. The impact of EM mutual coupling is omitted for simplicity.}
\end{figure*}

Once the operator's SVD is determined, power allocation can be carried out by a water-filling approach to achieve the CAPA capacity. However, deriving the SVD for a general radiation operator is mathematically demanding and is feasible only for specific cases, such as line-of-sight (LoS) far-field channels. For more general scenarios, determining the SVD requires solving a Fredholm integral equation of the second kind. This problem is often addressed by approximating the operator's kernel as a degenerate kernel, which reduces the task to solving a system of linear equations. Fourier series expansions are commonly used to approximate the kernel. 

Another critical property of compact operators is their ``almost'' finite-dimensional nature. Their singular values are countably infinite and exhibit a step-like pattern: remaining constant initially before rapidly declining to zero. The index of the singular value at this transition defines the number of effective DoFs (EDoFs). This property allows the degenerate kernel approximation to be limited to a finite number of terms corresponding to the EDoFs. {\figurename} {\ref{figSectionIV_Figure_1a}} illustrates the single-user capacity achieved by CAPAs, which is computed using a finite Fourier series expansion to approximate the radiation operator. The results show that CAPAs can achieve a higher capacity than SPDAs.

\begin{figure*}[!t]
\centering
    \subfigure[Diversity gain vs. multiplexing gain.]
    {
        \includegraphics[height=0.23\textwidth]{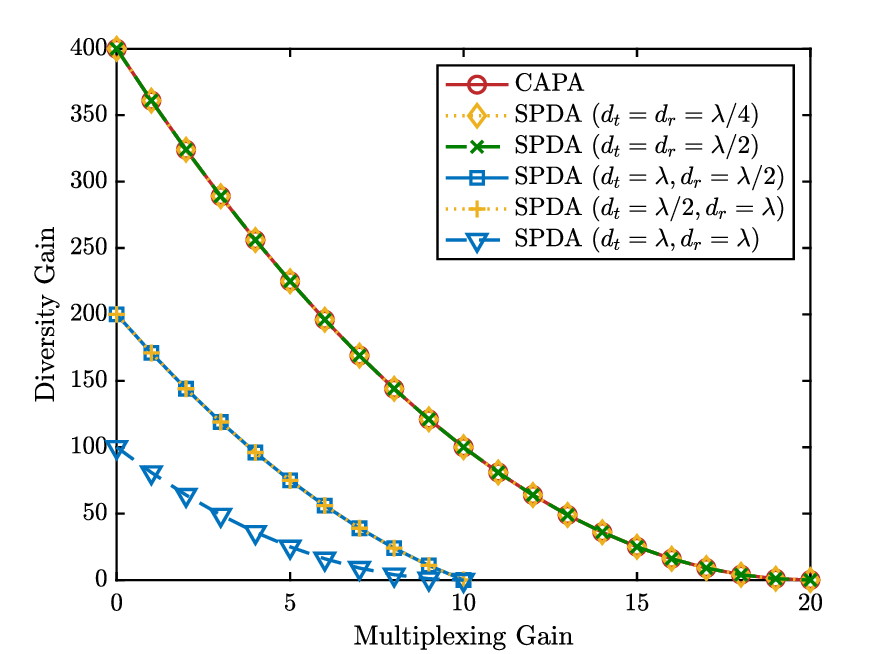}
	   \label{figSectionIV_Figure_2b}	
    }
   \subfigure[Ergodic channel capacity]
    {
        \includegraphics[height=0.23\textwidth]{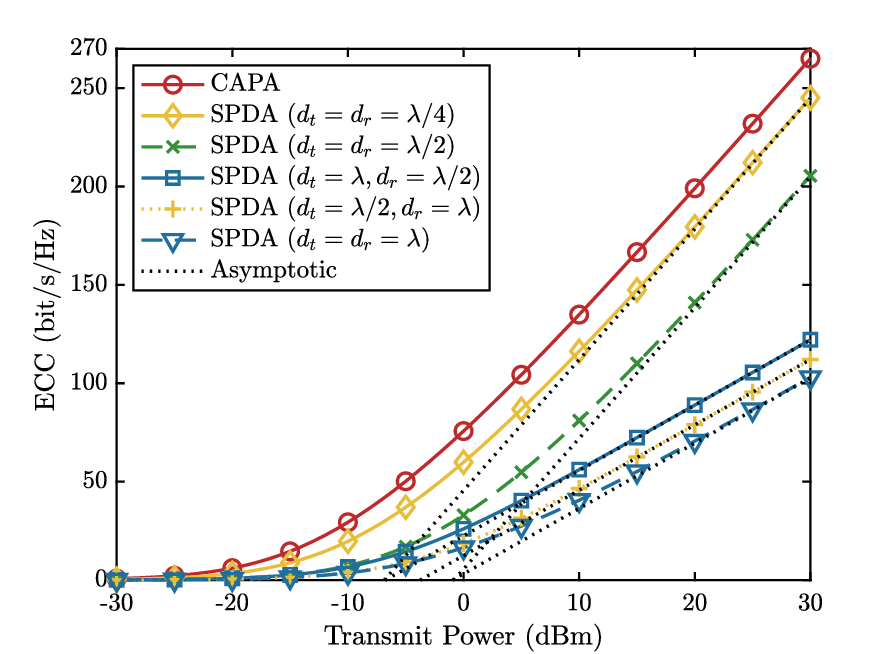}
	   \label{figSectionIV_Figure_2c}	
    }
\caption{The diversity and multiplexing gains and the ergodic channel capacity in a single-user fading channel between two CAPAs. The wavelength is denoted as $\lambda$. The parameters $d_{{\mathsf{t}}}$ and $d_{{\mathsf{r}}}$ represent the antenna spacing at the transmit SPDA and receive SPDA, respectively. The impact of EM mutual coupling is omitted for simplicity. Other simulation parameters can be found in \cite{ouyang2024diversity}.}
\vspace{-0.5cm}
\end{figure*}
\subsubsection{Multiuser Channel Capacity}
The capacity of multi-user channels becomes a multi-dimensional region defining the set of all rate vectors achievable simultaneously by all users. We consider the multiple access channel (MAC) and the broadcast channel (BC), where a CAPA serves multiple uplink and downlink users, each equipped with a single discrete antenna. 

In SPDA-based MACs, the capacity is achieved through successive interference cancellation (SIC) decoding, i.e., users' signals are successively decoded and subtracted from the received signal. During decoding, MMSE beamforming is applied to whiten interference and noise while maximizing the desired signal power. In BCs, the capacity is achieved using dirty-paper coding (DPC), a technique that enables pre-subtraction of interference known non-causally at the transmitter. When encoding signals for each user, MAC-BC duality is applied to design the beamformer that conveys the encoded symbols for the BC from its dual MAC \cite{el2011network}. The dual MAC is obtained by reversing the BC network while maintaining the same per-user channel gains as the original BC.

The multiuser capacity of CAPA-based MACs and BCs is also achieved using MMSE-SIC decoding and DPC-duality encoding. Recent findings in \cite{wang2024beamforming,ouyang2024performance} indicate that MMSE beamforming for CAPA-based MACs can be computed by solving a Fredholm integral equation with a separable integral kernel. This equation has a closed-form solution expressed as a weighted sum of each user's spatial response. Based on this approach, the multiuser capacity for CAPA-MACs under any SIC decoding order can be determined. The capacity of CAPA-BCs was explored using DPC-duality encoding for the two-user case in \cite{zhao2024continuous}. However, the MAC-BC duality in this study relies on heuristic assumptions and cannot be generalized to scenarios with more than two users. Intuitively, the capacity-achieving DPC beamformer for CAPA-BCs should also reside in the signal subspace spanned by users' spatial responses, which could help establish a more robust MAC-BC duality framework for CAPAs.

{\figurename} {\ref{figSectionIV_Figure_1b}} and {\figurename} {\ref{figSectionIV_Figure_1c}} illustrate the two-user capacity regions of CAPA-MACs and CAPA-BCs, respectively. In {\figurename} {\ref{figSectionIV_Figure_1b}}, the corner points represent different SIC orders, with the remaining region formed by time-sharing between these points. The BC capacity region in {\figurename} {\ref{figSectionIV_Figure_1c}} is the convex hull of its dual MAC capacity regions. These figures demonstrate that CAPAs achieve larger capacity regions than SPDAs in both MAC and BC scenarios. Additionally, increasing the aperture size further expands the capacity region.

\subsection{Diversity and Multiplexing Gains}
While the performance of a CAPA in a constant channel can be clearly described by its capacity, the performance in a fading channel can be more appropriately measured by outage probability (OP) and ergodic channel capacity (ECC). For CAPA-based fading channels, two crucial performance indicators derived from the OP and ECC are the multiplexing and diversity gains. For brevity, we consider a single-user channel between two CAPAs.

The diversity gain reflects the signal-to-noise ratio (SNR) exponent of the OP in high-SNR regimes. It measures the number of independently faded paths a transmitted symbol can travel through. The multiplexing gain quantifies how the data rate scales with SNR at high SNRs, in contrast to that for single-antenna channels. While the multiplexing gain captures improvements in data throughput, the diversity gain indicates the system's reliability. A tradeoff exists between these two gains, as achieving a higher multiplexing gain often reduces the SNR exponent of the OP. This tradeoff, known as the diversity-multiplexing tradeoff (DMT), reveals how system reliability diminishes when prioritizing data throughput and vice versa. In SPDA-based systems, the diversity gain, multiplexing gain, and DMT are limited by the number of discrete antennas, which can be analyzed using the random matrix theory.

As discussed in Section \ref{Section: Single-User Channel Capacity}, any physically realistic radiation operator is ``almost'' finite-dimensional with an ``almost finite'' rank, which is proven to be band-limited in the wavenumber (spatial-frequency) domain. Based on the Nyquist sampling theorem, the operator can be losslessly approximated by a weighted sum-product of two orthogonal bases, such as Fourier bases, with the weights derived from its wavenumber-domain representation. Consequently, the maximum achievable diversity or multiplexing gain is limited by the Nyquist number of the radiation operator, which depends on the CAPA's geometry and the surrounding scattering environment. Existing research has established that this Nyquist number equals the number of EDoFs of the radiation operator. This connection enables the application of random matrix theory to analyze the DMT of CAPAs using the wavenumber-domain sampling matrix. For instance, our previous work examined the DMT of two CAPAs under isotropic Rayleigh fading \cite{ouyang2024diversity}. The findings reveal that the EDoF of the radiation operator scales proportionally with the aperture size of the CAPAs and inversely with the wavelength.

For SPDAs with half-wavelength antenna spacing or less, the achievable DMT closely approaches that of CAPAs, as the number of antennas reaches the system's EDoFs \cite{sanguinetti2022wavenumber}. However, for SPDAs with antenna spacing larger than half the wavelength, CAPAs demonstrate a clear advantage in the DMT performance, as illustrated in {\figurename} {\ref{figSectionIV_Figure_2b}}. Additionally, due to the full utilization of spatial resources, CAPAs achieve higher array gains than SPDAs, resulting in higher ECCs, as illustrated in {\figurename} {\ref{figSectionIV_Figure_2c}}.

\section{Concluding Remarks and Future Research}
This article has provided a systematic introduction to the applications of CAPAs in wireless communications. We introduced three hardware prototypes of CAPAs, namely electronic, optical, and acoustic designs. Furthermore, we discussed three approaches for CAPA beamforming design: the discretization approach, the CoV approach, and the subspace approach. Numerical results demonstrated the low complexity and near-optimal performance of the CoV and subspace approaches compared to the discretization approach. We also analyzed the performance gains of CAPAs regarding channel capacity and DMT. This article is hoped to provide high-level guidance for future research in this area. The investigation of CAPAs is still in its infancy, and many open research problems remain. Some are outlined below:
\begin{itemize}
    \item {\textbf{Channel Estimation for CAPAs:} Given that the signal model for CAPA is integral-based and continuous, which corresponds to an infinite-dimensional channel, the conventional estimation methods, such as least squares or linear MMSE, are no longer applicable. However, channel estimation approaches that exploit structural properties and can reconstruct channels regardless of dimensionality remain promising, such as parametric estimation methods and beam training. Additionally, emerging machine learning techniques, such as physics-informed neural networks, are also promising for addressing integral-based channel estimation challenges.}
  \item \textbf{Wideband Transmission for CAPAs:} CAPAs are expected to operate in high-frequency bands, where wideband multi-carrier transmission will be a common feature. Extending existing matrix-based multi-carrier frameworks, such as orthogonal frequency-division multiplexing (OFDM), to CAPAs requires further investigation. Additionally, CAPA-enabled analog wideband beamforming introduces the issue of beam squint. Developing effective methods to mitigate beam squint is an important research direction.
  \item \textbf{Tri-polarized Beamforming for CAPAs:} Most existing research assumes CAPAs are uni-polarized, meaning the current distribution is aligned in the same polarization direction. However, practical CAPAs can be tri-polarized, where the current distribution at different locations within the CAPA varies in polarization. In this case, the current distribution evolves from a scalar function to a vector function, transforming the beamforming design into a more complex but interesting new research problem.      
\end{itemize}
\bibliographystyle{IEEEtran}
\bibliography{mybib}
\end{document}